\documentclass[12pt]{article}
\usepackage{graphicx}
\usepackage[margin=1.25in]{geometry}
\usepackage{url}
\usepackage{bm}
\usepackage{picture}
\usepackage{lineno}

\newcommand\pubnumber{FERMILAB--PUB--14--068--T\\ SLAC--PUB--15936}
\newcommand\pubdate{April, 2014}

\def\SLAC{SLAC,
    Stanford University, Menlo Park, California 94025 USA}
\def\doeack{\footnote{Work supported by the US Department of Energy,
                     contract DE--AC02--76SF00515.}}
\def\Cornell{Laboratory for Elementary-Particle Physics,
Cornell University, \\ Ithaca, New York 14853, USA}
\def\cornellack{\footnote{Work supported in part by the National 
Science Foundation.}}
\def\FNAL{Fermi National Accelerator Laboratory, Batavia, IL  60510  USA}
\def\fermiack{\footnote{Work supported by the US Department of Energy,  
contract DE--AC02--07CH11359. }}
\def\Title#1{\begin{center} {\Large #1 } \end{center}}
\def\Author#1{\begin{center}{ \sc #1} \end{center}}
\def\Address#1{\begin{center}{ \it #1} \end{center}}

\newcommand\pubblock{\rightline{\begin{tabular}{l} \pubnumber\\
         \pubdate \end{tabular}}}
\newenvironment{Abstract}{\begin{quotation} \begin{center}
                       ABSTRACT
     \end{center}\bigskip  }{\end{quotation}}

\def\Acknowledgements{\bigskip  \bigskip \begin{center} \begin{large}
             \bf ACKNOWLEDGEMENTS \end{large}\end{center}}

\def\beq{\begin{equation}}
\def\eeq#1{\label{#1}\end{equation}}
\def\eeqn{\end{equation}}

\newenvironment{Eqnarray}%
   {\arraycolsep 0.14em\begin{eqnarray}}{\end{eqnarray}}
\def\beqa{\begin{Eqnarray}}
\def\eeqa#1{\label{#1}\end{Eqnarray}}
\def\eeqan{\end{Eqnarray}}
\def\CR{\nonumber \\ }

\def\leqn#1{(\ref{#1})}

\def\overbar#1{\overline{#1}}

\let\bar=\overbar

\def\etal{{\it et al.}}

\def\lsim{\mathrel{\raise.3ex\hbox{$<$\kern-.75em\lower1ex\hbox{$\sim$}}}}
\def\gsim{\mathrel{\raise.3ex\hbox{$>$\kern-.75em\lower1ex\hbox{$\sim$}}}}

\def\half{\frac{1}{2}}

\def\del{\partial}
\def\Dslash{\not{\hbox{\kern-4pt $D$}}}
\def\dslash{\not{\hbox{\kern-2pt $\del$}}}

\def\msb{{\bar{\scriptsize M \kern -1pt S}}}

\def\drb{{\bar{\scriptsize D \kern -1pt R}}}

\makeatletter
\def\section{\@startsection{section}{0}{\z@}{5.5ex plus .5ex minus
 1.5ex}{2.3ex plus .2ex}{\large\bf}}
\def\subsection{\@startsection{subsection}{1}{\z@}{3.5ex plus .5ex minus
 1.5ex}{1.3ex plus .2ex}{\normalsize\bf}}
\def\subsubsection{\@startsection{subsubsection}{2}{\z@}{-3.5ex plus
-1ex minus  -.2ex}{2.3ex plus .2ex}{\normalsize\sl}}

\renewcommand{\@makecaption}[2]{%
   \vskip 10pt
   \setbox\@tempboxa\hbox{\small #1: #2}
   \ifdim \wd\@tempboxa >\hsize     
       \small #1: #2\par          
     \else                        
       \hbox to\hsize{\hfil\box\@tempboxa\hfil}
   \fi}

 \def\citenum#1{{\def\@cite##1##2{##1}\cite{#1}}}
 
\newcount\@tempcntc
\def\@citex[#1]#2{\if@filesw\immediate\write\@auxout{\string\citation{#2}}\fi
  \@tempcnta\z@\@tempcntb\m@ne\def\@citea{}\@cite{\@for\@citeb:=#2\do
    {\@ifundefined
       {b@\@citeb}{\@citeo\@tempcntb\m@ne\@citea\def\@citea{,}{\bf ?}\@warning
       {Citation `\@citeb' on page \thepage \space undefined}}%
    {\setbox\z@\hbox{\global\@tempcntc0\csname b@\@citeb\endcsname\relax}%
     \ifnum\@tempcntc=\z@ \@citeo\@tempcntb\m@ne
       \@citea\def\@citea{,}\hbox{\csname b@\@citeb\endcsname}%
     \else
      \advance\@tempcntb\@ne
      \ifnum\@tempcntb=\@tempcntc
      \else\advance\@tempcntb\m@ne\@citeo
      \@tempcnta\@tempcntc\@tempcntb\@tempcntc\fi\fi}}\@citeo}{#1}}
\def\@citeo{\ifnum\@tempcnta>\@tempcntb\else\@citea\def\@citea{,}%
  \ifnum\@tempcnta=\@tempcntb\the\@tempcnta\else
  {\advance\@tempcnta\@ne\ifnum\@tempcnta=\@tempcntb \else\def\@citea{--}\fi
    \advance\@tempcnta\m@ne\the\@tempcnta\@citea\the\@tempcntb}\fi\fi}
\makeatother

\newcommand{\MSbar}{$\overline{\mathrm{MS}}$}

\begin{document}
\begin{titlepage}
\pubblock

\vfill
\Title{Expected Precision of Higgs Boson Partial Widths within the
  Standard Model}
\vfill
\Author{G.~Peter Lepage$^a$\cornellack, Paul B. Mackenzie$^b$\fermiack, and 
Michael E. Peskin$^c$\doeack}
\Address{$a$. \Cornell \\ $b$. \FNAL \\ $c$. \SLAC}

\vfill
\begin{Abstract}
We discuss the sources of uncertainty in calculations of 
the partial widths of the Higgs boson within the Standard
Model.   The uncertainties come from two sources: the 
truncation of perturbation theory and the uncertainties in 
input parameters.   We review the current status of perturbative
calculations and note that these are  already reaching the 
parts-per-mil level of accuracy for the   major decay modes.
   The main sources of 
uncertainty will then come from  the parametric dependences on 
$\alpha_s$, $m_b$, and $m_c$.  Knowledge of these parameters
is systematically improvable through lattice gauge theory 
calculations.  We estimate the precision that lattice QCD
will achieve in the next decade and the corresponding
precision of the Standard Model predictions for Higgs boson
partial widths.
\end{Abstract}
\vfill
\end{titlepage}

\thispagestyle{empty}
\tableofcontents








\newpage
\pagenumbering{arabic}
\def\thefootnote{\fnsymbol{footnote}}
\setcounter{footnote}{0}

\section{Introduction}

After the discovery of the Higgs boson by the ATLAS and CMS
collaborations~\cite{ATLAS,CMS}, much attention has been given to the 
measurement of the properties of this particle.   In principle, 
accurate measurements of the Higgs properties can tell us
whether the corresponding Higgs field is the sole source of mass
for quarks, leptons, and gauge bosons, and whether there 
are new particles that also receive mass from this field.   The 
report \cite{Higgsworking} for the 2013 Snowmass Community 
Summer Study reviews the 
current status of measurements of the Higgs boson couplings
and projections of the capabilities of future collider programs.

However accurately the couplings of the Higgs boson are 
measured, though, these measurements are  useful only if 
combined with comparably accurate predictions from the 
Standard Model (SM).   New physics associated with the 
Higgs boson appears as deviations from the SM
predictions.  The report \cite{Higgsworking} gives
many examples of new physics effects that alter the 
Higgs boson couplings at the few-percent level. The
discovery of these effects  will  require both the measurements and
theory of these couplings
to have uncertainties below the percent level.   If deviations cannot be discerned because of 
intrinsic uncertainty in the the theoretical  predictions, the goal of
the program of precision measurements on the Higgs
boson will be frustrated.    

In particular, the proposed experiments at the 
International Linear Collider have demonstrated the 
capability of measuring individual Higgs boson couplings
in a model-independent way
to the level of parts per mil~\cite{ILCHiggs,myILCHiggs}. 
This seems to us an important goal, but it is 
only important if the
SM predictions for Higgs boson couplings can be 
given with similar accuracy.

Currently, the partial widths of the Higgs boson within
the SM are generally agreed to be predicted to accuracies of a few percent.
This situation is summarized in the work of the 
LHC Higgs Cross Section Working Group~\cite{HiggsBRs,HiggsBRstoo}
and in a recent paper by Almeida, Lee, Pokorski, and
Wells~\cite{WellsHiggsBRs}.   This latter paper 
presents a significant challenge:
\begin{quote}
``... the SM uncertainty in computing $B(H\to b\bar b)$ is 
presently 3.1\% (sum of absolute values of all errors) and expected to 
not get better than 2.2\%, with most of that coming from the 
uncertainty of the bottom Yukawa coupling determination ... Thus,
without a higher-order calculation to substantially reduce this error,
any new physics contribution to the $b\bar b$ branching fraction
that is not at least a factor of two or three 
larger than 2\% cannot be discerned.  Thus, a deviation of 
at least 5\% is  required of detectable new physics.''~\cite{WellsHiggsBRs}
\end{quote}

We agree with the general conclusions of
\cite{HiggsBRs,HiggsBRstoo,WellsHiggsBRs}
as far as the current situation is concerned, and
we will often refer to these useful papers in our
analysis below.  However, we believe that the quote in the
previous paragraph, which applies the current uncertainties
to experiments that will be done a decade from now and
 draws pessimistic conclusions, is seriously misleading.
Most importantly,
it underestimates the power of lattice QCD  to 
give us precision knowledge of the $b$ quark mass and
of its renormalization to the Higgs boson mass scale.
We will argue here that the SM predictions for the 
Higgs boson partial width to $b\bar b$, and for the 
other dominant decay modes, will be improved to the 
parts-per-mil level on a time scale that matches the 
needs of the High-Luminosity LHC and ILC experimental
programs.

This paper is organized as follows:  In Section 2, we 
develop basic  notation for our study of Higgs partial
width uncertainties.  In Section 3, we review perturbative
computations of the partial widths to the dominant SM
decay modes and the uncertainties that they imply. Our conclusion 
is that it is within the current state of the art to reduce the 
uncertainties from missing terms in perturbation theory to the 
parts-per-mil level.  For many of the Higgs boson partial widths, this 
is already achieved.

 In Section 4, we discuss the determination 
of the most important
input parameters --- $\alpha_s$
and the $b$ and $c$ quark masses ---
from lattice gauge theory.  Data from lattice QCD simulations can be used to 
determine the QCD parameters in several different ways. The most 
straightforward method to describe is to  compute the spectrum of 
heavy-quark mesons, adjust the parameters of the lattice action to 
fit the measurements, and then convert these parameters to 
a continuum definition (for example, \MSbar\ subtraction).  This   
method is typically limited by the accuracy of existing lattice QCD
perturbation theory calculations.  An alternative and more promising 
method is to use lattice simulations to predict continuum quantities
such as QCD sum rules that can be  readily interpreted
using 
continuum QCD calculations.  It is worth noting that 
almost all of the 
highest-precision determinations of $\alpha_s$ and many of the 
highest-precision determinations of $b$ and $c$ mass reported by the
Particle Data Group~\cite{PDG}  use this  strategy.
In Section 4, we illustrate this 
approach with the lattice calculation of the moments of $b$ and $c$
quark pseudoscalar current correlation functions.  These correlation 
functions were used  in \cite{McN} to provide measurements 
of $\alpha_s$, $m_b$ and $m_c$  with accuracy at the current state
of the art.   Using toy Monte 
Carlo calculations, we estimate how much the uncertainties computed
in \cite{McN} could be decreased 
over the next decade using  the increased computer resources
that should become available over this time.
Section 5 gives our conclusions.

\section{Structure of Higgs boson partial widths}

In this paper, we will quote uncertainties using the 
Higgs boson partial widths.  Our common coin will be 
the relative theoretical  uncertainty $\delta_A$ on the extracted coupling of
the Higgs boson to $A\bar A$, which 
we will take uniformly to be $\half$ of the uncertainty on the 
the corresponding partial width.   
\beq
      \delta_A  =   \half {\Delta \Gamma(h\to A\bar A)\over \Gamma(h\to
        A\bar A)} .
\eeq{defDelta}
 For definiteness, 
we set the Higgs boson mass to $m_h = 126.0$~GeV
throughout this paper.

More generally in this paper, we will use the symbol $\Delta$ to denote an absolute
uncertainty on a measurable quantity, and $\delta$ to denote the
relative uncertainty,
\beq
       \delta X =    {\Delta X\over X} \ .
\eeq{defsmdelta}
In this notation,  $\delta_A = \half \delta \Gamma(h\to A\bar A)$. 

There are two contributions to the $\delta_A$.  The first
 is
the {\it theoretical error} due to the fact that the perturbation
theory
is computed only up to a certain order.  As we will see, theoretical 
errors for the $\delta_A$ are, in almost all cases, already at the 
few parts-per-mil level.   The second is the {\it parametric error}
due  to the uncertainties of needed input parameters.  These 
parametric errors will have most of our attention in the paper.

In
\cite{HiggsBRs} and \cite{WellsHiggsBRs},
uncertainties are quoted for the prediction of Higgs branching
ratios.  We prefer to work with partial widths, because these
are more primitive objects.   Branching ratios are composites
that depend on all of the partial widths, through
\beq
    BR(h\to A\bar A) = {\Gamma(h\to A\bar A)\over \sum_C \Gamma(h\to
    C\bar C)}  \ , 
\eeq{BRcomp}
where the sum over $C$ runs over all decay modes.
This can potentially lead to some confusion. 
 For example, in Table IV of \cite{HiggsBRs}, the
authors quote an uncertainty of 2\% in the branching ratios 
$BR(h\to \tau^+\tau^-)$ and $BR(h \to WW^*)$ for a 120 GeV Higgs 
boson due to parametric 
dependence on the $b$ quark mass.  This comes entirely from the 
dependence on $\Gamma(h\to b\bar b)$ in the denominator of 
\leqn{BRcomp}  and has nothing to do with the Higgs couplings to 
$\tau^+\tau^-$ or $WW$.  This impression is rectified in the 
presentation in Table 1 of \cite{HiggsBRstoo}. 
 We  note that the complete program 
of Higgs boson measurements  planned for the ILC allows the 
absolutely normalized partial widths to be extracted in a
model-independent
way~\cite{ILCHiggs}. 

A  Higgs boson partial width typically has the structure
\beq
    \Gamma(h\to A\bar A) =   {G_F\over \sqrt{2}} {m_h  m_A^2 \over
      4\pi} \cdot {\cal F} 
\eeq{GHstruct}
where ${\cal F}$ is a scalar function of coupling constants and mass ratios.
The 
factor $m_A^2$ arises from the fact that the Higgs coupling to 
$A\bar A$ is proportional to $m_A$.  It is often the case that the dominant 
contribution to the parametric uncertainty in $\Gamma(H\to A\bar A)$
comes from this term. 

The contributions to $\delta_A$ 
from the first two terms of the  prefactor are
\beqa
     \delta_A &=&  \half \delta G_F \oplus \half \delta m_h
     \CR
          & = &  (3\times 10^{-7}) \oplus (1.2\times 10^{-4})  \ ,
\eeqa{DeltaApref}
where the first term uses the  current uncertainty~\cite{PDG} and 
 the second term assumes a Higgs boson mass measurement with an 
uncertainty of 30~MeV, as expected at the ILC~\cite{ILCHiggs}.  The
dependence
on the Higgs mass is larger in the $h\to gg, \gamma\gamma, \gamma Z$
partial widths, which are proportional to $m_h^3$.  However, it is
only 
non-negligible for the  partial
widths to $WW^*$ and $ZZ^*$, which depend strongly on the available 
phase space.   The uncertainty in  \leqn{DeltaApref}  
coming from  $m_A$ depends on the particle species in question.   For 
$\tau$, $W$, and $Z$, there are well-defined on-shell values
which 
are known quite accurately~\cite{PDG}:
\beq
   \delta m_\tau = 9\times 10^{-5} \qquad
\delta m_W =  1.9\times10^{-4}      \qquad   \delta m_Z =
2.3\times 10^{-5} \ .
\eeq{Deltams}
These estimates give the impression, which is also correct in the
complete
theory, that the uncertainties in Higgs couplings due to the
uncertainties in 
these input parameters are negligible.

For quark and gluon final states, the situation is quite different.
Well-defined on-shell states are not theoretically accessible, and so 
we must rely on  QCD perturbation theory, which potentially
brings in sizable parametric uncertainties. 

 QCD  perturbation theory
 is best behaved if one
evaluates
Higgs partial widths using the 
\MSbar\ mass evaluated at the Higgs boson mass scale.
However, the masses of the quarks are usually
quoted at a much lower scale, either as the perturbative pole masses or
as the \MSbar\ masses at 
scale  near the quark threshold.   The conversion of these mass values 
to \MSbar\ masses at $m_h$ is often a dominant uncertainty in 
the prediction of the Higgs boson couplings.

QCD sum rules measure off-shell quark masses at momenta of the order
of $2m_Q$.    The Higgs boson couplings are obtained most accurately
by directly extrapolating these values to $m_h$.  We will use the 
\MSbar\ masses  $m_b$(10.0 GeV) and $m_c$(3.0 GeV) as our
basic inputs.  The conversion of  a
mass value at $2m_Q$ to a pole mass brings in a substantial
QCD uncertainty, 
and there is an additional uncertainty in converting the pole mass
back
to an \MSbar\ value at $m_h$.   The papers
\cite{HiggsBRs,HiggsBRstoo,WellsHiggsBRs} use the pole masses as
inputs.
This leads to a stronger dependence on the input mass and $\alpha_s$
values than what we quote below and, consequently, an overestimate
of the uncertainty.

 The QCD theory of the evolution of 
mass parameters is nicely reviewed by Che\-tyr\-kin, K\"uhn, and 
Steinhauser in \cite{RunDec}, with a computer code {\tt RunDec}
 implementing
their prescriptions with terms up to NNNLO also provided.  
 The uncertainty 
in the conversion from low scale masses to  $m_Q(m_h)$ due to the 
truncation of perturbation theory is small:  for example, 
the NNNLO terms 
in the series give  a relative correction of $0.8\times 10^{-4}$.  In 
the following, and in our later discussion of QCD effects,
\beq
            a(\mu)  =   {\alpha_s^{\overbar{MS}}(\mu)\over \pi}
\eeq{adefin}

Using the notation of \cite{RunDec},
 the parametric uncertainties in $m_b(m_h)$ are proportional to the 
derivatives
\beqa
  {m_b(10) \over m_b(m_h)} { d m_b(m_h)\over d m_b(10)} &=& 1  \CR
 {\alpha_s(m_Z)\over m_b(m_h)} { d m_b(m_h)\over d \alpha_s(m_Z)} &=&
     a(m_Z)\cdot  {\gamma_m(a(m_h)) -
  \gamma_m(a(10))\over
   \beta(a(m_Z))}   =  -0.38  \ .
\eeqa{mbconvert}
The numerical values are computed using 5-flavor running and the 
current PDG value $\alpha_s(m_Z) = 0.1185$.   Note that the derivative in the first line is
reduced by taking a fixed renormalization point of 10.0~GeV for $m_b$ rather than 
one that depends on $m_b$.   If we took $m_b(m_b)$ as a reference,
this coefficient would be 1.19; for the pole mass, this coefficient is
1.28. 
We will also need the conversion factor
\beq
{\alpha_s(m_Z)\over \alpha_s(m_h)}{ d \alpha_s(m_h)\over d
\alpha_s(m_Z)} = {\alpha_s(m_Z)\over \alpha_s(m_h)}  {\beta(a(m_h))\over
\beta(a(m_Z))}   = 0.95 \ .
\eeq{alphaconvert}

For the input variable $m_c$(3), we need to take into account 
4-flavor running between $m_b$ and the reference point.  The bulk
of the effect is accounted in 
\beqa
 {\alpha_s(m_Z)\over m_c(m_h)} { d m_c(m_h)\over d \alpha_s(m_Z)} & =&
  \CR
& & \hskip -1.3in    a(m_Z)\cdot  {\gamma_m(a(m_h)) -
  \gamma_m(a(m_b)) + \gamma^{(4)}_m(a^{(4)}(m_b)) -
 \gamma^{(4)}_m(a^{(4)}(3))\over
   \beta(a(m_Z))}   =  -0.91   \ .
\eeqa{mcconvert}
There is also a small dependence of 
$m_c(m_h)$ on the position of the matching point $m_b$,  given 
approximately by
\beq
  {m_b(m_b)\over m_c(m_h)} {d m_c(m_h)\over d m_b(m_b)} = 
     2 (\gamma_m(a(m_b)) - \gamma^{(4)}_m(a^{(4)}(m_b))) =  0.004 \ .
\eeq{mcgap}
Finally, there are tiny discontinuities between the 5-flavor
and 4-flavor formulae that sightly change the dependences given in 
these two equations.   We quote the final result in \leqn{mbmcsumm}.

In all, we find that the term $m_Q^2$ in \leqn{GHstruct}, for the 
cases of $Q = b$ or $c$,  gives a
contribution to the uncertainty from the parametric dependence
on quark masses and on $\alpha_s$.  This dependence is 
given by 
\beqa
     \delta m_b(m_h) &=& 1.0 \cdot \delta m_b(10)
    \oplus (-0.38) \cdot \delta \alpha_s(m_Z) \CR
     \delta m_c(m_h)  &=& 1.0 \cdot \delta m_c(3) \oplus(- 0.90)
       \cdot \delta \alpha_s(m_Z)
  \oplus ( 0.006)\cdot \delta m_b(10) \ .
\eeqa{mbmcsumm}
The coefficients in this expression are of order 1, so it is already
clear that very accurate
values
of the parameters on the right-hand side are needed to predict Higgs
partial widths to part-per-mil accuracy.

\section{Perturbation theory for Higgs boson partial widths}

With the orientation given in the previous section, we now review the
status of perturbative computations of the partial width for the major
decay modes of the SM Higgs boson.  A detailed  overview of SM Higgs 
decay modes is given in  Djouadi's review paper~\cite{Djouadi}. 
That discussion has been updated 
in \cite{HiggsBRstoo,WellsHiggsBRs}.  In particular, Table~3 of
\cite{WellsHiggsBRs} gives the parametric dependence of the
predictions 
for the full set of input parameters.  However, since we are using a 
different scheme of inputs,we must revisit the
dependence on the most important parameters $m_b$, $m_c$, and
$\alpha_s$.

\subsection{$h\to b\bar b$}

The corrections to the partial width $\Gamma(h\to b\bar b)$ can be 
grouped as (i) QCD corrections to the correlation function of scalar
currents $b\bar b$, (ii) additional QCD corrections involving flavor
singlet intermediate states, (iii)  electroweak corrections and mixed
QCD/electroweak corrections.   All terms are proportional to 
$m_b^2(m_h)$.  The dominant corrections are of the type (i).  

The corrections of type (i) are known to ${\cal O}(\alpha_s^4)$
through
a very impressive calculation of Baikov, Chetyrkin, and
K\"uhn~\cite{BCK}.   They evaluate to 
\beqa
\tilde R &=& 1 + 5.667 a + 29.15 a^2 + 41.76 a^3 -
825.7 a^4 \CR
  &=& 1 +  0.2037 +  0.0377 +  0.0019 -0.0013  \ ,
\eeqa{BCKresult}
so that the series seems to be converging, with a residual error at the 
part-per-mil level in $\delta_b$~\cite{Wang}.  The parametric dependence of
\leqn{BCKresult} on $\alpha_s$ is obtained as
\beq
         \frac{\alpha_s(m_h)}{\tilde R}
      {d\over d\alpha_s(m_h)} \tilde R = 0.22
\eeq{BCKparam}
This must be combined with the dependence of the prefactor given
in \leqn{mbmcsumm}.

 The
corrections (ii) begin in ${\cal O}(a^2)$, are known to ${\cal
  O}(a^3)$,
and are less than 1\%
corrections to $\delta\Gamma_b$~\cite{Kwiat,ChStein}. 

 For the corrections of type (iii), 
the complete ${\cal O}(\alpha)$ result is
known~\cite{Bardin,Kniehlbb,Dabelstein}, but at the 2-loop level
only the leading terms of  ${\cal O}( \alpha a
m_t^2/m_h^2)$~\cite{Kwiattb,KniehlSpira}
and  ${\cal O}( \alpha^2 
m_t^4/m_h^4)$~\cite{Buttenschorn} have been computed.
Numerically, these three terms are, respectively,
\beq
{\delta\Gamma} =  0.3\% - 0.02\% + 0.05\%
\eeq{numbb}

Thus, the theoretical understanding of this decay is already 
close to the part-per-mil level in $\delta_b$.   The parametric
dependence on the most important parameters is
\beq
  \delta_b = 1. \cdot  \delta m_b(10) \oplus (-  0.28)  \cdot  \delta
  \alpha_s(m_Z) \ .
\eeq{finalforb}
In~\cite{myILCHiggs}, it was estimated that the $hb\bar b$ coupling
would be measured to  0.3\% at the ILC in its late stages.

\subsection{$h\to c\bar c$}

The theoretical calculation of the partial width 
$\Gamma(h\to c\bar c)$ is essentially
 the same as that for $h \to b\bar b$.  In
particular, 
the qualitative picture that the theory is close to part-per-mil
accuracy continues to hold.   The parametric uncertainty, combining
\leqn{mbmcsumm} and \leqn{BCKparam}, is 
\beq
  \delta_c = 1. \cdot  \delta m_c(3) \oplus( - 0.80) \cdot  \delta
  \alpha_s(m_Z) \ .
\eeq{finalforc}
In~\cite{myILCHiggs}, it was estimated that the $hc\bar c$ coupling
 would be measured to  0.7\% at the ILC in its late stages.

\subsection{$h\to\tau^+\tau^-$}

The theoretical calculation of the partial width 
$\Gamma(h\to\tau^+\tau^-)$ is very similar to 
that for $h \to b\bar b$, except that there are no 
QCD corrections except for universal ones.  We see no
issue here in obtaining a precise SM prediction.
In~\cite{myILCHiggs}, it was estimated that the $h\tau^+\tau^-$
coupling
would be measured to  0.7\% at the ILC in its late stages.

\subsection{$h\to gg$}

The theoretical prediction for the partial width $\Gamma(h\to gg)$ 
begins in ${\cal O}(a^2)$.  The series of QCD corrections has been 
computed to ${\cal O}(a^4)$ by Schreck and Steinhauser~\cite{SStein}, 
with each term given as a series in $\tau = m_h^2/4m_t^2$. Baikov and 
Chetyrkin~\cite{BC} and Moch and Vogt~\cite{MV} have also obtained 
the leading term at ${\cal O}(a^5)$.  If $\Gamma_0$ is the
leading-order 
result for $m_t \gg m_h$, the series evaluates to 
\beqa
{\Gamma\over \Gamma_0} &=& 1.0671 + 19.306 a + 172.76  a^2 + 467.68 a^3\CR
   &=&   1.0671 + 0.6942 + 0.2234 + 0.0217 
\eeqa{ggseries}
 The parametric dependence of
\leqn{ggseries} on $\alpha_s$ is obtained as
\beq
   {  \alpha_s(m_h)\over \Gamma } {d\over d\alpha_s(m_h)} \Gamma =  2.6 
\eeq{SSparam}
There is also an electroweak correction of $+5\%$, known only to the 
leading order (which is already ${\cal O}(\alpha a^2)$), computed by 
Actis, Passarino, Sturm, and Uccirati~\cite{Actis}. At the 1\%
  level, some final states produced by the $hgg$ coupling contain
  $b\bar b$ due to gluon splitting.  It should be clarified through
  simulation
to what extent these final states will be classified by the
experiments
as $h\to b\bar b$ rather
than
$h\to gg$ decays.

We find that the uncertainty from theory in the prediction of the $hgg$ coupling is 
now at the 1\% level. This situation is improvable, although computing 
additional orders of perturbation theory will be challenging.
The important parametric dependence of the SM prediction is
\beq
  \delta_g = 1.2 \cdot \delta \alpha_s(m_Z)  \ .
\eeq{finalforg}
In~\cite{myILCHiggs}, it was estimated that the $hgg$
coupling
would be measured to  0.6\% at the ILC in its late stages.

\subsection{$h\to \gamma\gamma$}

For $\Gamma(h\to \gamma\gamma)$, the leading term
is ${\cal O}(\alpha^2)$.  The electroweak correction
of order  ${\cal O}(\alpha^2)$.  has been computed by 
Passarino, Sturm, and Uccirati~\cite{Pass}, and the QCD
corrections of  ${\cal O}(\alpha\alpha_s^2)$ and 
 ${\cal O}(\alpha\alpha_s^3)$ have 
been computed by Maierh\"ofer and Marquard~\cite{MandM}. 
The relative sizes of the corrections are, respectively, 
\beq
      -  1.6\%  + 1.8\% + 0.08\% \ . 
\eeq{finalforgamma}
The uncertainty in the prediction of the $h\gamma\gamma$
coupling is, then, at the parts-per-mil level, and there is no
significant parametric uncertainty. 
In~\cite{myILCHiggs}, it was estimated that the $h\gamma\gamma$
coupling
would eventually 
be measured to  0.8\% using a combination of LHC and ILC
results.

\subsection{$h\to WW^*$, $h\to ZZ^*$}

The situation for the decays $h\to WW^*$ and $h\to ZZ^*$ is somewhat
more complicated, and beyond the scope of this paper to explain in
full.   The decay involves color-singlet particles in leading order,
so  the radiative corrections are at the percent level.  The complete
${\cal O}(\alpha_s)$ and ${\cal O}(\alpha)$ corrections have been 
computed by Bredenstein, Denner, Dittmaier, and Weber~\cite{Breden}.
These authors find corrections of, for example,  1\% for $h\to 
e^+e^- \mu^+\mu^-$, 3\% for $h \to \nu_e e^+ \mu^-\bar\nu_\mu$, 
7\% for $h \to \nu_e e^+ q\bar q$,  and 10\% for $h\to q\bar q q \bar
q$. Quite consistently, the difference between the full radiative
corrections
and those of the Improved Born Approximation (IBA), in which the two 
off-shell vector bosons are treated separately, is 1\%. Additional 
corrections to the IBA are known, including corrections of 
${\cal O}(\alpha a  m_t^2/m_h^2)$, ${\cal O}(\alpha a^2  m_t^2/m_h^2)$, 
and ${\cal O}(\alpha^2 m_t^4/m_h^4)$.  These corrections are reviewed
in \cite{Veretin}; they bring the calculation of this approximation to 
the part-per-mil level.  A full 2-loop analysis without the IBA
approximation
 will be more
difficult.

These partial widths have no important parametric uncertainty due to 
$\alpha_s$ or $m_b$, but they do depend strongly on the mass of the 
Higgs boson.  From \cite{WellsHiggsBRs} (based on \cite{Breden}),
\beq
     \delta_W = 6.9 \cdot \delta m_h\ , \quad  
     \delta_Z = 7.7\cdot \delta m_h \ .
\eeq{WZparam}
That is, a 
measurement of the Higgs boson mass to 30~MeV precision would lead 
to a 0.2\% theoretical uncertainty in these partial widths. 

In~\cite{myILCHiggs}, it was estimated that the $hWW$ and $hZZ$
couplings would each be measured to 0.2\% at the ILC in its late
stages.
It seems within the state of the art for theory to match this level 
of accuracy, though it will be a challenge.

\section{Improved parameters from lattice QCD}

One of the implications of the previous section is that the SM
predictions
for several of the Higgs boson partial widths depend strongly 
on $\alpha_s$, $m_b$, $m_c$, and $\alpha_s$.   The current values of 
these parameters are 
\beqa
    \alpha_s(m_Z) &=& 0.1185 \pm 0.0006 \qquad\quad\ (\pm 0.5\%)\CR
     m_b(10) &=& 3.617 \pm 0.025~\mbox{GeV}  \qquad \quad  (\pm 0.7\%) \CR
     m_c(3)   & = & 0.986 \pm 0.006~\mbox{GeV} \qquad  (\pm 0.6\%)
 \eeqa{currentparams}
The first of line of \leqn{currentparams} is  the current Particle Data Group
value~\cite{PDG}.   The second and third lines are lattice gauge
theory determinations, from \cite{McN}.   These are consistent with
the PDG averages, with a slightly larger error for $m_b$ and  a
somewhat smaller error for $m_c$.   Using the values in
\leqn{currentparams},
assuming that the errors are uncorrelated and that it is
  correct to combine errors in quadrature, we find the parametric components of the 
uncertainty in Higgs coupling predictions to be 
\beq
     \delta_b =  0.7\%  \ , \quad  \delta_c =  0.7\% \ ,   \quad   \delta_g =  0.6\% \ . 
\eeq{currentdelta}
This is already quite impressive accuracy, but the future program of
precision Higgs measurements will require that we do better.  

In the rest of this section we will describe how one uses lattice QCD (LQCD) to 
extract the $\overline{\mathrm{MS}}$~coupling and masses. We will
illustrate these ideas with a simple example, and use that example
to explore what improvements will be possible over the next decade.
Finally we will briefly survey other approaches from LQCD that
are likely to contribute over that period.

\subsection{Lattice QCD}
In LQCD, continuous 
space and time are replaced by a discrete mesh 
of lattice sites with a lattice spacing~$a$ that is typically of
order 0.15\,fm or less. The path integral
describing QCD becomes an ordinary multidimensional integral
in this approximation, with the lattice functioning as 
the ultraviolet regulator. Lattice simulations integrate the 
path integral numerically, using Monte Carlo methods, to 
obtain Monte Carlo estimates for vacuum expectation values
of a wide variety of operators from which physics is extracted.

Having chosen a value for the bare coupling, the first step
in an LQCD simulation is to tune the bare quark masses and the 
lattice spacing to values that reproduce physical results from
the real world. 
The masses are typically adjusted to reproduce 
experimental results for particular, well-measured hadron masses: 
for example, $m_\pi$, $m_K$, $m_{\eta_c}$, and $m_{\eta_b}$.   The
lattice spacing is set using some other well-measured
quantity, such as the pion decay constant~$f_\pi$. Once one
has tuned
these parameters, renormalized matrix elements 
from a LQCD simulation will
agree with the corresponding matrix elements from
continuum  QCD up to errors
of $\mathcal{O}(a^2)$. Simulations are generally performed at 
multiple values of~$a^2$ and results extrapolated to~$a=0$.

The quark masses and the QCD coupling constant are specified quite accurately 
by the tuning process, but they are defined for the lattice 
regulator, not the \MSbar\ regulator typically used in 
continuum calculations. In principle, bare
lattice masses and couplings can be converted
into~$\overline{\mathrm{MS}}$ quantities using perturbation 
theory\---\,see, for example, \cite{Lee:2013mla} 
and~\cite{Davies:2008sw}. In practice, however,
the precision of this approach has been limited by the
difficulty involved in calculating the conversion formulae,
which require high-order perturbative calculations using
the (very complicated) lattice regulator. 

A different approach that has proven quite successful is to use
lattice simulations to generate nonperturbative values for
renormalized short-distance quantities, such as matrix
elements of current-current correlators at short distances. 
Renormalized quantities are regulator independent, and 
therefore values obtained from LQCD simulations can be 
analyzed using ordinary continuum~\MSbar\ 
perturbation theory once they have been extrapolated 
to zero lattice spacing. Such analyses can be used
to extract values for the~\MSbar\ 
coupling and masses in the same way that values are 
extracted from experimental data.

\subsection{An example}
One quantity that can be used to compute 
all three of our important input parameters $m_b$, $m_c$ and~$\alpha_s$
is the current-current correlator
\beq
  G(t) \equiv a^3 \sum_{\mathbf{x}} m_{0Q}^2
  \langle 0 | j_{5Q}(\mathbf{x},t)\,j_{5Q}(0,0) | 0 \rangle
\eeq{Gdefin}
where $j_{5Q} \equiv \overline{\psi}_Q\gamma_5\psi_Q$
is the pseudoscalar density for a heavy quark~$Q$
(either~$c$
or~$b$)~\cite{McN}. 
This correlator is a close relative of  the vector current-current correlators, which may  be
obtained from $e^+e^-$ heavy-quark production. The data for
  these
  vector correlators, analyzed with continuum perturbation theory,
  currently provide the most precise determinations of the heavy
  quark masses~\cite{Chetyrkin:2009fv}.
We consider only the connected correlator, where both 
currents are on the same quark line, since this simplifies both
the lattice simulation and the continuum analysis. The factors
of the LQCD bare quark mass~$m_{0Q}$ in \leqn{Gdefin} make $G(t)$ ultraviolet 
finite. Then the  lattice and continuum versions are 
equal up to finite-lattice spacing corrections:
\beq
  G_\mathrm{cont}(t) = G_\mathrm{lat}(t) + \mathcal{O}(a^2).
\eeq{close}

Low-$n$ moments of $G(t)$,
\beq
  G_{2n} \equiv a\sum_t t^{2n} G(t) 
  = (-1)^n \frac{\partial^{2n}}{\partial E^{2n}}G(E=0),
\eeq{moments}
are perturbative, 
since the energy $E=0$ at which they are evaluated is far below
the threshold $E\approx2m_Q$
for heavy quarks.
Moments with $2n\ge4$ are ultraviolet finite.  Consequently,
 in perturbation theory, we obtain
\beq
  G_{2n} = 
  \frac{g_{2n}(\alpha_{\overline{\mathrm{MS}}}(\mu))}{m_Q(\mu)^{2n-4}}
\eeq{Gtwon}
where $\mu$ is the renormalization scale and
  $g_{2n}(\alpha_{\overline{\mathrm{MS}}})$
is a perturbation series known through third order for $2n = 4$, 6, 8,
and 10~\cite{Chetyrkin:2006xg,Boughezal:2006px,Maier:2008he,Maier:2009fz,Kiyo:2009gb}.  The scale $\mu$ should
be taken close to $2m_Q$ to avoid large logarithms. One then adjusts the 
values of the
\MSbar\ coupling and quark masses so that the 
(continuum) perturbative expressions agree
with the nonperturbative values for the moments $G_{2n}$ generated by LQCD. 

A detailed LQCD analysis of correlator moments is given in~\cite{McN}.
 It replaces the moments $G_{2n}$ by reduced moments~$R_{2n}$ in order 
to suppress systematic errors caused by the simulation; statistical
errors
 from the Monte Carlo are insignificant. The values obtained for 
$m_c(3\,\mathrm{GeV})$ and $\alpha_{\overline{\mathrm{MS}}}(M_Z)$ are 
accurate to 0.6\%, while $m_b(10\,\mathrm{GeV})$ is accurate to 0.7\%.
 The dominant source of uncertainty in the first two quantities is the
 lack 
of $4^\mathrm{th}$-order perturbation theory. The dominant error in
 $m_b(10\,\mathrm{GeV})$ comes from the finite lattice spacing, which
 matters more for the $b$~quark because of its larger mass.

An LQCD simulation offers several advantages over experiment as
a source for nonperturbative results, beyond the obvious fact
that it is easier to instrument a simulation than an experiment.
For example, we can produce results not only for $m_Q=m_c$ and
$m_Q=m_b$, but also for several quark masses in between~$m_c$
and~$m_b$. This allows us to vary the value of 
$\alpha_{\overline{\mathrm{MS}}}(\mu)$,  since $\mu \sim 2m_Q$,
and therefore to use the simulation data to estimate and bound
perturbative corrections beyond third order. The result is a much
more reliable estimate of the perturbative errors than comes from the
standard procedure of replacing $\mu$ by~$\mu/2$ and~$2\mu$. 
Varying the quark mass also allows us to probe and fit the 
leading nonperturbative behavior, from the gluon condensate.
The Operator Product Expansion implies that
\begin{equation}
  G_{2n} = G_{2n}^\mathrm{short-distance}
  \left\{
  1 + d_{2n}(\alpha_{\overline{\mathrm{MS}}}) 
  \frac{\langle \alpha_s G^2/\pi\rangle}{(2m_Q)^4} + \cdots
  \right\}
\end{equation}
In practice the condensate correction turns out
to be negligible compared to other
uncertainties, because it is suppressed by $1/(2m_Q)^4$.

LQCD simulations also allow us for the first time to determine
ratios of quark masses nonperturbatively~\cite{McN,Davies:2009ih}.
These ratios, which can be determined quite accurately,
provide a highly non-trivial check on values obtained from 
perturbative methods, and can be used to leverage a precise
determination of one mass into precise determinations of other 
masses.

\begin{figure}
  \centering
  \includegraphics[width=0.7\textwidth]{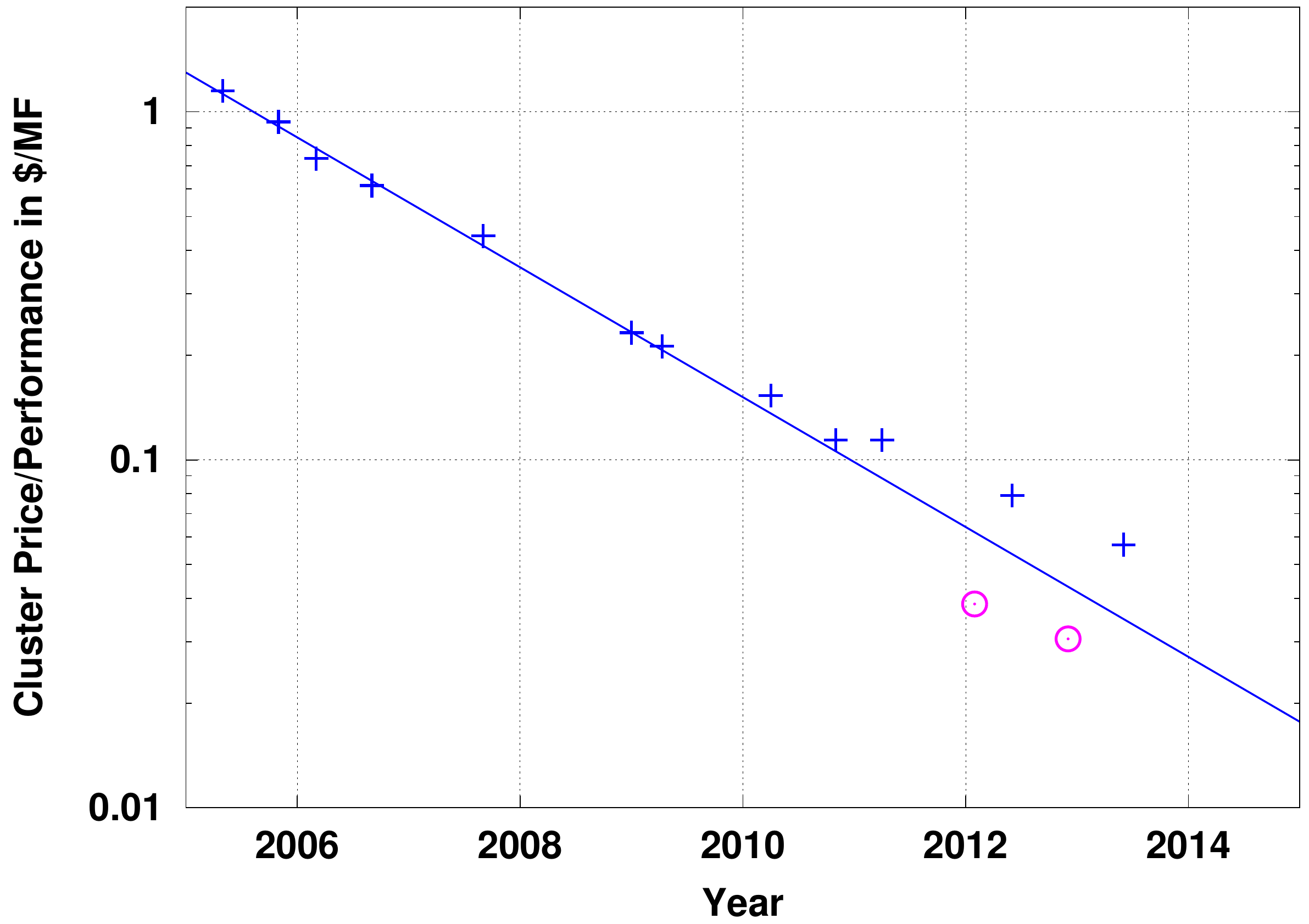}
  \caption{  \label{Fig:cpucost}
 Measured cost per megaflop of lattice QCD computing on the
  USQCD cluster facilities at Fermilab and Jefferson Lab,  plotted versus year.
  The exponentially improving price/performance of conventional cluster
   hardware (blue crosses) that was observed through
  2011 has fallen off somewhat in the last few years.
  This has been mitigated by the introduction, where possible, of GPU-accelerated clusters
  (magenta circles) for lattice calculations.}
\end{figure}

\subsection{Projections}

The LQCD analysis described above has yielded the heavy quark
masses and the QCD coupling constant with precisions that are already
below 1\%.
A detailed study of those results shows that the most important
limiting factors are the lack of higher-order perturbation 
theory ($4^\mathrm{th}$-order) and the finite lattice
spacing~\cite{McN}.
Both sources of error can potentially be reduced.  It seems
feasible, given time, to compute the next term in perturbation theory
for the correlator moment, bringing these to $4^\mathrm{th}$~order in 
$\alpha_s$. 
The extent to which 
the lattice spacing can be reduced depends upon further
reductions in the cost of computing. Figure~\ref{Fig:cpucost}
shows that computing costs have fallen by roughly a factor of~100
since~2005 at the USQCD facilities at Fermilab and Jefferson Lab. Similar reductions
are expected over the next 10--15~years. Simulation costs
  scale 
roughly as $1/a^6$, and so we expect
  that the smallest lattice space used for this simulation could be
reduced by about a factor of two, from 0.045\,fm used in \cite{McN} to
0.023\,fm.

There are more issues to face, beyond securing adequate
hardware, if one wishes to achieve per-mil accuracy in LQCD simulations.
Some, like the inclusion of isospin violation and electromagnetic corrections,
 we expect will be straightforward.
It is possible that some may prove greater challenges.
For example,
topological structure in the 
gauge field develops more slowly in simulations at smaller 
lattice spacing, which may pose problems at very small 
lattice spacing. 
Such issues might require new ideas, but
the tremendous advances in LQCD over the past decade make us
optimistic that any new obstacles 
will be overcome during the next 10--15~years.

The existing lattice analysis can be used to predict the impact
of  these improvements in the order of perturbation 
theory and the size of the lattice spacing 
 on the precision with which we can determine 
the coupling constant and masses from the correlator 
moments. The current analysis compares results from 
multiple lattice spacings in order
to determine the dependence of LQCD results on 
the lattice spacing. This allows us to 
extrapolate existing results to smaller lattice spacings.
By adding realistic noise to these extrapolations,
we create synthetic data for smaller lattice
spacings that can be combined with 
existing LQCD data in a new analysis of the 
masses and coupling. The results tell us
the extent to which smaller lattice spacings 
reduce errors on the masses and coupling. The impact
of higher-order perturbation theory is also
easily evaluated by adding fake $4^\mathrm{th}$-order
terms to the perturbation theory.

We have gone through this exercise starting from the 
analysis in~\cite{McN}. Our results are presented in 
Table~\ref{projected}. This table shows the percent errors we 
expect in the masses and coupling from the correlator analysis
under various scenarios for improvements: PT denotes the 
effect of computing QCD perturbation theory through
$4^\mathrm{th}$~order.  LS denotes the effect of decreasing
the lattice spacing to 0.03\,fm.  LS$^2$ denotes the effect 
of using lattices with 0.03\,fm and 0.023\,fm lattice spacing.
We recall that the stage LS$^2$ corresponds to 
an increase in computing power by about a factor of 100.
ST denotes the effect of improving the statistics by a factor
of 100.   We also 
show percent errors for the Higgs couplings to
$b\bar b$, $c\bar c$, and~$gg$, accounting 
for correlations among the errors in the determination
of the parameters. The last line of the table gives, for 
comparison,
the experimental uncertainties in the Higgs boson 
couplings expected after the ILC
measurements~\cite{myILCHiggs}.

\begin{table*}
    \begin{tabular}{r|ccc|ccc}
         & $\delta m_b(10) $ 
         & $\delta \alpha_s(m_Z)$ 
         & $\delta m_c(3)$ & 
        $\delta_{b}$ 
        & $\delta_{c}$ & $\delta_{g}$ \\ \hline\hline
       current errors \cite{McN}
            & 0.70 & 0.63 & 0.61 & 0.77 & 0.89 & 0.78
        \\ & & & & & & \\
     + PT
            & 0.69 & 0.40 & 0.34 & 0.74 & 0.57 & 0.49
        \\ 
     +  LS       
            & 0.30 & 0.53 & 0.53 & 0.38 & 0.74 & 0.65
        \\
     +  LS$^2$ 
            & 0.14 & 0.35 & 0.53 & 0.20 & 0.65 & 0.43
        \\ & & & & & & \\
     +  PT + LS
            & 0.28 & 0.17 & 0.21 & 0.30 & 0.27 & 0.21
        \\
   +  PT + LS$^2$
            & 0.12 & 0.14 & 0.20 & 0.13 & 0.24 & 0.17
        \\
    +   PT + LS$^2$ + ST
            & 0.09 & 0.08 & 0.20 & 0.10 & 0.22 & 0.09
        \\ & & & & & & \\
        ILC goal 
            &  &  & & 0.30 & 0.70 & 0.60
        \end{tabular}
\caption{\label{projected}Projected fractional errors, in percent, for the 
  $\overline{\mathrm{MS}}$~QCD coupling and heavy quark masses under different
  scenarios for improved analyses. The improvements considered are:
  PT - addition of $4^\mathrm{th}$~order QCD perturbation theory, LS, LS$^2$ -
  reduction of the lattice spacing to 0.03\,fm and to 0.023\,fm; ST -
  increasing the statistics of the simulation by a factor of 100. The last three
  columns 
convert the errors in input parameters into errors on Higgs couplings, 
taking account of correlations.  The bottom line gives the
target
values of these errors suggested by the projections for the ILC
measurement
accuracies.}
\end{table*}

We find that reducing the lattice spacing to 0.023\,fm
is sufficient to bring parametric errors for
the Higgs couplings below the errors expected from
the full ILC. Adding $4^\mathrm{th}$-order perturbation
theory reduces the parametric errors
further, to about half of the expected ILC~errors. Adding 
statistics gives a relatively small further reduction in the errors.

These error 
estimates are likely conservative because they assume that
there will be no further innovation in LQCD simulation methods. 
There already are many alternative lattice methods for
extracting the QCD coupling from LQCD simulations: see, for 
example,~\cite{Davies:2008sw,Shintani:2010ph,Bazavov:2012ka,Jansen:2011vv,Fritzsch:2012wq}. 
None of these methods involve 
heavy quark masses directly and so none have correlations between
$\alpha_s$  and heavy quark masses.   Small lattice spacings are
important for an accurate $b$~mass because of $(am_b)^2$~errors;
these can be avoided completely by using effective field theories
such as NRQCD~\cite{Lee:2013mla}
or the Fermilab formalism~\cite{ElKhadra:1996mp} for $b$-quark dynamics
in correlators, rather than (highly corrected) relativistic
actions.  Other renormalized 
lattice matrix elements, such
as off-shell expectation values of $m_Q \overline{\psi}\gamma_5\psi$,
can be used to compute masses~\cite{Martinelli:1994ty}. 
There are many ideas that are likely
to come into play over the next decade or so.

\section{Conclusions}

In this paper, we have surveyed the current status of the
uncertainties in the Standard Model calculations of the Higgs boson
partial widths.  We have shown that the current theory of these 
partial widths is already accurate to better than 1\%.  We have also 
seen that both the perturbation theory and the parametric inputs 
to this theory can be improved, to yield predictions at an accuracy 
beyond even the high level expected for the experiments at the ILC.
Lattice gauge theory has a crucial role to play in improving the 
determination of the most important input parameters.

There is much work to be done in the next decade to realize
the program we have outlined.  But the result will be that these 
calculations, combined with the results of precision experiments,
will offer a powerful probe into the mysteries of the Higgs boson.

\Acknowledgements

The general idea for a connection between Higgs physics and lattice
gauge
theory calculation arose in discussions for the Snowmass 2013 Energy
Frontier reports.  We are especially grateful to Sally Dawson, Heather Logan,
Laura   Reina, 
Chris  Tully, and Doreen Wackeroth for providing the impetus for this
study.
MEP thanks Ayres Freitas for an introduction to the literature on
precision
Higgs calculations and Sven Heinemeyer, Michael Spira, and James Wells for
illuminating discussions.
We thank Don Holmgren for the data and plot in Fig.~\ref{Fig:cpucost}.
 
The work of GPL was supported by the National Science Foundation. 
GPL would also like to thank
the Department of Applied Mathematics and Theoretical Physics
at Cambridge University for their hospitality while this work 
was in progress.
The work of PBM was supported by the US Department of Energy,  
Contract No. DE-AC02-07CH11359.
 The work of  MEP was supported by 
the U.S. Department 
of Energy under contract DE--AC02--76SF00515.


\begin{thebibliography}{99}



\bibitem{ATLAS}
 G.~Aad {\it et al.}  [ATLAS Collaboration],
  Phys.\ Lett.\ B {\bf 716}, 1 (2012)
  [arXiv:1207.7214 [hep-ex]].

\bibitem{CMS} 
S.~Chatrchyan {\it et al.}  [CMS Collaboration],
  Phys.\ Lett.\ B {\bf 716}, 30 (2012)
  [arXiv:1207.7235 [hep-ex]].
 
\bibitem{Higgsworking}
S. Dawson, \etal, Higgs boson working group report for Snowmass 2013, 
arXiv:1310.8361.


\bibitem{ILCHiggs}
 D.~M.~Asner, {\it et al.},
  arXiv:1310.0763 [hep-ph].


\bibitem{myILCHiggs}
  M.~E.~Peskin,
  arXiv:1312.4974 [hep-ph].
 

\bibitem{HiggsBRs}
 A.~Denner, S.~Heinemeyer, I.~Puljak, D.~Rebuzzi and M.~Spira,
  Eur.\ Phys.\ J.\ C {\bf 71}, 1753 (2011)
  [arXiv:1107.5909 [hep-ph]].

\bibitem{HiggsBRstoo}
 S. Heinemeyer {\it et al.}  [LHC Higgs Cross Section Working Group Collaboration],
  arXiv:1307.1347 [hep-ph].


\bibitem{WellsHiggsBRs}
L.~G.~Almeida, S.~J.~Lee, S.~Pokorski and J.~D.~Wells,
  arXiv:1311.6721 [hep-ph].

\bibitem{PDG}
J. Beringer et al. (Particle Data Group), Phys. Rev. {\bf D86}, 010001
(2012),
updated at \url{http://pdg.lbl.gov}.

\bibitem{McN}
C.~McNeile, C.~T.~H.~Davies, E.~Follana, K.~Hornbostel and G.~P.~Lepage,
  Phys.\ Rev.\ D {\bf 82}, 034512 (2010)
  [arXiv:1004.4285 [hep-lat]].


\bibitem{RunDec}
 K.~G.~Chetyrkin, J.~H.~Kuhn and M.~Steinhauser,
  Comput.\ Phys.\ Commun.\  {\bf 133}, 43 (2000)
  [hep-ph/0004189];
B.~Schmidt and M.~Steinhauser,
  Comput.\ Phys.\ Commun.\  {\bf 183}, 1845 (2012)
  [arXiv:1201.6149 [hep-ph]].

\bibitem{Djouadi}
 A.~Djouadi,
  Phys.\ Rept.\  {\bf 457}, 1 (2008)
  [hep-ph/0503172].


\bibitem{BCK}
 P.~A.~Baikov, K.~G.~Chetyrkin and J.~H.~Kuhn,
  Phys.\ Rev.\ Lett.\  {\bf 96}, 012003 (2006)
  [hep-ph/0511063].

\bibitem{Wang}
In this paper, we estimate the error from truncation of QCD
perturbation theory as the value of the last computed term.  This is
more conservative than the usual procedure of estimating this error
from $\mu$ scale dependence.  The scale dependence of perturbative
expressions for the  Higgs boson partial width is analyzed, and
systematically improved, in 
 S.~-Q.~Wang, X.~-G.~Wu, X.~-C.~Zheng, J.~-M.~Shen and Q.~-L.~Zhang,
  arXiv:1308.6364 [hep-ph].  

\bibitem{Kwiat}
 K.~G.~Chetyrkin and A.~Kwiatkowski,
  Nucl.\ Phys.\ B {\bf 461}, 3 (1996)
  [hep-ph/9505358].

\bibitem{ChStein}
 K.~G.~Chetyrkin and M.~Steinhauser,
  Phys.\ Lett.\ B {\bf 408}, 320 (1997)
  [hep-ph/9706462].

\bibitem{Bardin}
 D.~Y.~.Bardin, B.~M.~Vilensky and P.~K.~.Khristova,
  Sov.\ J.\ Nucl.\ Phys.\  {\bf 53}, 152 (1991)
  [Yad.\ Fiz.\  {\bf 53}, 240 (1991)].

\bibitem{Kniehlbb}
 B.~A.~Kniehl,
  Nucl.\ Phys.\ B {\bf 376}, 3 (1992).

\bibitem{Dabelstein}
 A.~Dabelstein and W.~Hollik,
  Z.\ Phys.\ C {\bf 53}, 507 (1992).
 
\bibitem{Kwiattb}
  A.~Kwiatkowski and M.~Steinhauser,
  Phys.\ Lett.\ B {\bf 338}, 66 (1994)
  [Erratum-ibid.\ B {\bf 342}, 455 (1995)]
  [hep-ph/9405308].

\bibitem{KniehlSpira}
 B.~A.~Kniehl and M.~Spira,
  Nucl.\ Phys.\ B {\bf 432}, 39 (1994)
  [hep-ph/9410319].

\bibitem{Buttenschorn}
 M.~Butenschoen, F.~Fugel and B.~A.~Kniehl,
  Phys.\ Rev.\ Lett.\  {\bf 98}, 071602 (2007)
  [hep-ph/0612184],
  Nucl.\ Phys.\ B {\bf 772}, 25 (2007)
  [hep-ph/0702215 [HEP-PH]].

\bibitem{SStein}
 M.~Schreck and M.~Steinhauser,
  Phys.\ Lett.\ B {\bf 655}, 148 (2007)
  [arXiv:0708.0916 [hep-ph]].


\bibitem{BC}
 P.~A.~Baikov and K.~G.~Chetyrkin,
  Phys.\ Rev.\ Lett.\  {\bf 97}, 061803 (2006)
  [hep-ph/0604194].
 
\bibitem{MV}
 S.~Moch and A.~Vogt,
  Phys.\ Lett.\ B {\bf 659}, 290 (2008)
  [arXiv:0709.3899 [hep-ph]].

\bibitem{Actis}
  S.~Actis, G.~Passarino, C.~Sturm and S.~Uccirati,
  Nucl.\ Phys.\ B {\bf 811}, 182 (2009)
  [arXiv:0809.3667 [hep-ph]].

\bibitem{Pass}
 G.~Passarino, C.~Sturm and S.~Uccirati,
  Phys.\ Lett.\ B {\bf 655}, 298 (2007)
  [arXiv:0707.1401 [hep-ph]].
 
\bibitem{MandM}
 P.~Maierh\"ofer and P.~Marquard,
  Phys.\ Lett.\ B {\bf 721}, 131 (2013)
  [arXiv:1212.6233 [hep-ph]].
 
\bibitem{Breden}
 A.~Bredenstein, A.~Denner, S.~Dittmaier and M.~M.~Weber,
  Phys.\ Rev.\ D {\bf 74}, 013004 (2006)
  [hep-ph/0604011],
  JHEP {\bf 0702}, 080 (2007)
  [hep-ph/0611234].


\bibitem{Veretin}
 B.~A.~Kniehl and O.~L.~Veretin,
  Phys.\ Rev.\ D {\bf 86}, 053007 (2012)
  [arXiv:1206.7110 [hep-ph]].

\bibitem{Lee:2013mla} 
  A.~J.~Lee {\it et al.}  [HPQCD Collaboration],
  Phys.\ Rev.\ D {\bf 87}, no. 7, 074018 (2013)
  [arXiv:1302.3739 [hep-lat]].

\bibitem{Davies:2008sw} 
  C.~T.~H.~Davies {\it et al.}  [HPQCD Collaboration],
  Phys.\ Rev.\ D {\bf 78}, 114507 (2008)
  [arXiv:0807.1687 [hep-lat]]. See, specifically, results
  for $\alpha_\mathrm{lat}/W_{11}$.

\bibitem{Chetyrkin:2009fv} 
  K.~G.~Chetyrkin, J.~H.~Kuhn, A.~Maier, P.~Maierhofer, P.~Marquard, M.~Steinhauser and C.~Sturm,
  Phys.\ Rev.\ D {\bf 80}, 074010 (2009)
  [arXiv:0907.2110 [hep-ph]].


\bibitem{Chetyrkin:2006xg}
 K.~G.~Chetyrkin, J.~H.~Kuhn and C.~Sturm,
   Eur.\ Phys.\ J.\  C {\bf 48}, 107 (2006)
 [arXiv:hep-ph/0604234].


\bibitem{Boughezal:2006px}
 R.~Boughezal, M.~Czakon and T.~Schutzmeier,
  Phys.\ Rev.\  D {\bf 74}, 074006 (2006)
 [arXiv:hep-ph/0605023].

\bibitem{Maier:2008he}
  A.~Maier, P.~Maierhofer and P.~Marqaurd,
      Phys.\ Lett.\  B {\bf 669}, 88 (2008)
  [arXiv:0806.3405 [hep-ph]].

\bibitem{Maier:2009fz}
  A.~Maier, P.~Maierhofer, P.~Marquard and A.~V.~Smirnov,
    Nucl.\ Phys.\  B {\bf 824}, 1 (2010)
  [arXiv:0907.2117 [hep-ph]].

\bibitem{Kiyo:2009gb}
  Y.~Kiyo, A.~Maier, P.~Maierhofer and P.~Marquard,
    Nucl.\ Phys.\  B {\bf 823}, 269 (2009)
  [arXiv:0907.2120 [hep-ph]].




\bibitem{Davies:2009ih} 
  C.~T.~H.~Davies, C.~McNeile, K.~Y.~Wong, E.~Follana, R.~Horgan, K.~Hornbostel, G.~P.~Lepage and J.~Shigemitsu {\it et al.},
  Phys.\ Rev.\ Lett.\  {\bf 104}, 132003 (2010)
  [arXiv:0910.3102 [hep-ph]].

\bibitem{Shintani:2010ph} 
  E.~Shintani, S.~Aoki, H.~Fukaya, S.~Hashimoto, T.~Kaneko, T.~Onogi and N.~Yamada,
  Phys.\ Rev.\ D {\bf 82}, 074505 (2010)
  [arXiv:1002.0371 [hep-lat]].

\bibitem{Bazavov:2012ka} 
  A.~Bazavov, N.~Brambilla, X.~Garcia i Tormo, P.~Petreczky, J.~Soto and A.~Vairo,
  Phys.\ Rev.\ D {\bf 86}, 114031 (2012)
  [arXiv:1205.6155 [hep-ph]].

\bibitem{Jansen:2011vv} 
  K.~Jansen {\it et al.}  [ETM Collaboration],
  JHEP {\bf 1201}, 025 (2012)
  [arXiv:1110.6859 [hep-ph]].

\bibitem{Fritzsch:2012wq} 
  P.~Fritzsch, F.~Knechtli, B.~Leder, M.~Marinkovic, S.~Schaefer,
  R.~Sommer, and F.~Virotta [ALPHA Collaboration],
  Nucl.\ Phys.\ B {\bf 865}, 397 (2012)
  [arXiv:1205.5380 [hep-lat]].

\bibitem{ElKhadra:1996mp} 
  A.~X.~El-Khadra, A.~S.~Kronfeld and P.~B.~Mackenzie,
  Phys.\ Rev.\ D {\bf 55}, 3933 (1997)
  [hep-lat/9604004].

\bibitem{Martinelli:1994ty} 
  G.~Martinelli, C.~Pittori, C.~T.~Sachrajda, M.~Testa and A.~Vladikas,
  Nucl.\ Phys.\ B {\bf 445}, 81 (1995)
  [hep-lat/9411010].
 

\end{thebibliography}
\end{document}